\documentclass[3p,times]{elsarticle}

\usepackage{ecrc}

\volume{00}

\firstpage{1}

\journalname{Nuclear Physics A}

\runauth{Z.-W. Liu et al.}

\jid{NUPHA}

\jnltitlelogo{Nuclear Physics A}

\CopyrightLine{2012}{Published by Elsevier Ltd.}

\usepackage{amssymb}
\usepackage[figuresright]{rotating}
\usepackage{subfigure}
\usepackage{multirow}
\usepackage{amsmath}

\def\Tr{{\rm Tr}}

\allowdisplaybreaks

\begin{document}

\begin{frontmatter}

\dochead{}

\title{Pseudoscalar Goldstone Bosons Scattering off Charmed Baryons with Chiral Perturbation Theory}

\author[a]{Zhan-Wei Liu\corref{spk}}\cortext[spk]{Speaker}\ead{liuzhanwei@pku.edu.cn}
\author[a]{Shi-Lin Zhu}\ead{zhusl@pku.edu.cn}

\address[a]{Department of Physics and State Key Laboratory of Nuclear Physics and Technology, Peking University, Beijing 100871, China}

\begin{abstract}
We have systematically calculated the pseudoscalar Goldstone boson
and charmed baryon scattering lengths to the third order with
heavy baryon chiral perturbation theory. The scattering lengths
can reveal the interaction of $K \Lambda_c$, $K \Sigma_c$, $\pi
\Omega_c$ etc. For each channel, we take into account of the
interaction between the external particle and every possible
charmed baryon in the triplet, sextet, and excited sextet. We use
dimensional regularization and modified minimal subtraction to get
rid of the divergences in the loop-diagram corrections. We notice
that the convergence of chiral expansion for most of the channels
becomes better after we let the low energy constants absorb the
analytic contributions of the loop-diagram corrections and keep
the nonanalytic terms only.
\end{abstract}


\end{frontmatter}

\section{Introduction}\label{secInt}

There has been a growing interest in the properties of charmed
hadrons after Belle, BABAR, CDF and other collaborations observed
a series of interesting new states
\cite{Aubert2003,Krokovny2003,Aaltonen2009}. Much attention is
paid to the recently observed charmed hadrons
\cite{Wong2004,Kolomeitsev2004,Gamermann2007,Dai2008,Liu2009b,Liu2009a,Romanets2012}.
One may also wonder whether there exist some undiscovered new
states. Certainly, the interaction of charmed hadrons is important
to understand the underlying structure of these states.

The light pseudoscalar meson and charmed baryon scattering can
reflect the interaction between them. We will use the heavy baryon
chiral perturbation theory (HB$\chi$PT) to study the scattering
lengths between the light pseudoscalar meson and charmed baryon in
the triplet, sextet, and excited sextet to the third order. The
scattering lengths are expanded as a power series in $\epsilon$
with explicit power counting, where $\epsilon$ is the small
momentum of the light pseudoscalar mesons or the residual momentum
of the charmed baryons. The mass difference between charmed
baryons is also counted as $O(\epsilon^1)$.

The scattering length $a_{\phi B}$ is related to the $T$-matrix
$T_{\phi B}$ at the threshold by $T_{\phi
B}=4\pi(1+m_\phi/m_B)a_{\phi B}$. On the one hand, the positive
scattering length indicates that there exists attractive
interaction in these channels, which might lead to the loosely
bound molecular states. On the other hand, one can extract the
scattering length with the observed molecular states to test the
theory or determine the unknown low-energy constants
\cite{Baru2004,Meissner2012}.

This paper is organized as follows: in Sec. \ref{secLag}, we list
the HB$\chi$PT Lagrangians of the pseudoscalar mesons and charmed
baryons. In Sec. \ref{secSct}, we show the expressions of the
scattering lengths. In Sec. \ref{secRsl}, we discuss the numerical
results.

\section{Lagrangians}\label{secLag}

The leading scattering length is at $O(\epsilon^1)$, and receives
the tree-level contributions from Weinberg-Tomozawa terms of the
leading Lagrangian only. The scattering length at $O(\epsilon^2)$
receives the tree-level contributions from the Lagrangian at
$O(\epsilon^2)$. In addition to the tree-level contributions by
the Lagrangian at $O(\epsilon^3)$, the scattering length at
$O(\epsilon^3)$ also receives the one-loop corrections from the
leading Lagrangians.

The leading HB$\chi$PT Lagrangians read
\begin{eqnarray}
    {\mathcal L}_{\phi \phi}^{(2)}&=&f^2 \Tr\left( u_\mu u^\mu+\frac{\chi_+}{4} \right), \label{eqLPhiPhi} \\
    {\mathcal L}^{(1)}_{B\phi}&=&\frac12 \Tr[\bar B_{\bar 3} i v\cdot D B_{\bar 3}]
                            +\Tr[\bar B_{6} (i v\cdot D-\delta_2) B_{6}]
                            -\Tr[\bar B_{6}^* (i v\cdot D-\delta_3) B_{6}^*]\nonumber\\&&
                            +2 g_1 \Tr(\bar B_6 S\cdot u B_6)
                            +2 g_2 \Tr(\bar B_6 S\cdot u B_{\bar 3}+{\rm H.c.})
                            + g_3 \Tr(\bar B_{6 \mu}^* u^\mu B_{6}+{\rm H.c.})\nonumber\\&&
                            + g_4 \Tr(\bar B_{6 \mu}^* u^\mu B_{\bar 3}+{\rm H.c.})
                            +2 g_5 \Tr(\bar B_{6}^* S\cdot u B_{6}^*)
                            +2 g_6 \Tr(\bar B_{\bar 3} S\cdot u B_{\bar 3}), \label{eqLBPhiOne}
\end{eqnarray}
where $v_\mu=(1, \vec 0)$ is the velocity of a slowly moving
charmed baryon, $S_\mu$ is the spin matrix, and $\delta_i$ is the
mass difference between charmed baryons
\begin{equation}
  \delta_1=M_{B_6^*}-M_{B_6}=67 {~\rm MeV}, \quad
  \delta_2=M_{B_6}-M_{B_{\bar3}}=127 {~\rm MeV}, \quad
  \delta_3=M_{B_6^*}-M_{B_{\bar3}}=194 {~\rm MeV}.  \label{eqMssDif}
\end{equation}
$B_{\bar 3}$, $B_{6}$, and $B_{6}^*$ represent the charmed
triplet, sextet, and excited sextet, respectively. The notations
about the light pseudoscalar $\phi$ mesons are
\begin{equation}
\Gamma_\mu = \frac{i}{2} [\xi^\dagger, \partial_\mu\xi],\quad
u_\mu=\frac{i}{2} \{\xi^\dagger, \partial_\mu \xi\},\quad \xi
=\exp(i \frac{\phi}{2f}), \quad \chi_\pm
=\xi^\dagger\chi\xi^\dagger\pm\xi\chi\xi,\quad
\chi=\mathrm{diag}(m_\pi^2,\, m_\pi^2,\, 2m_K^2-m_\pi^2).
\label{equxiDef}
\end{equation}
$iD^\mu B_{ab}=i \partial^\mu B_{ab}+\Gamma_a^{\mu~d}
B_{db}+\Gamma_b^{\mu~d} B_{ad}$, which generates the
Weinberg-Tomozawa $BB\phi\phi$ vertices. The terms proportional to
$g_i$ in $ {\mathcal L}^{(1)}_{B\phi}$ generate the axial coupling
$BB\phi$ vertices.

The Lagrangians at $O(\epsilon^2)$ and $O(\epsilon^3)$ contain the
counter terms and recoil terms. The counter terms are constructed
based on the chiral symmetry, Lorentz invariance, C and P
invariance, and so on, while the recoil terms are derived by the
leading relativistic Lagrangians. They generate the
$BB\phi\phi$-like vertices. The concrete form of the Lagrangians
can be found in Ref. \cite{Liu2012b}.

\section{Scattering Lengths}\label{secSct}

\begin{figure}[htb]
\hspace{\stretch{0.5}}
 \subfigure[no axial vertices]{
 \scalebox{0.8}{\includegraphics{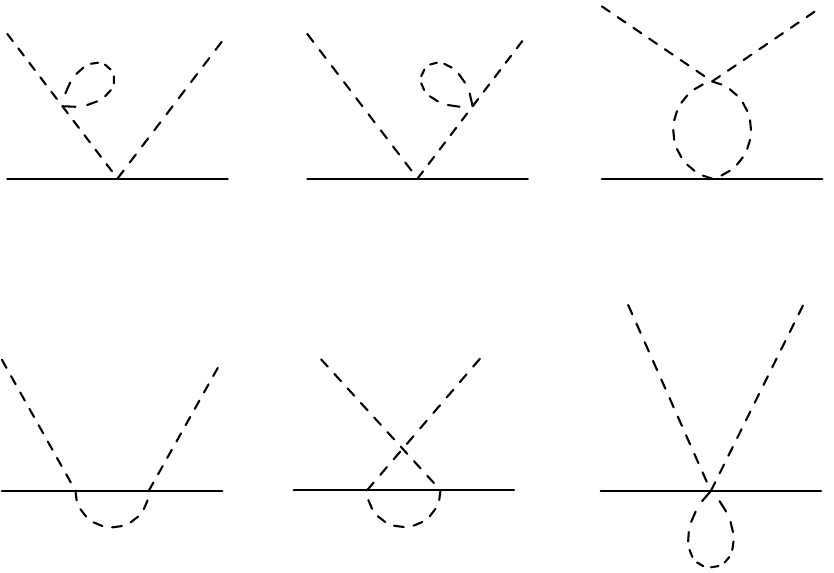}} \label{fig1a}
 }
 \hspace{\stretch{1}}
 \subfigure[two axial vertices]{
 \scalebox{0.8}{\includegraphics{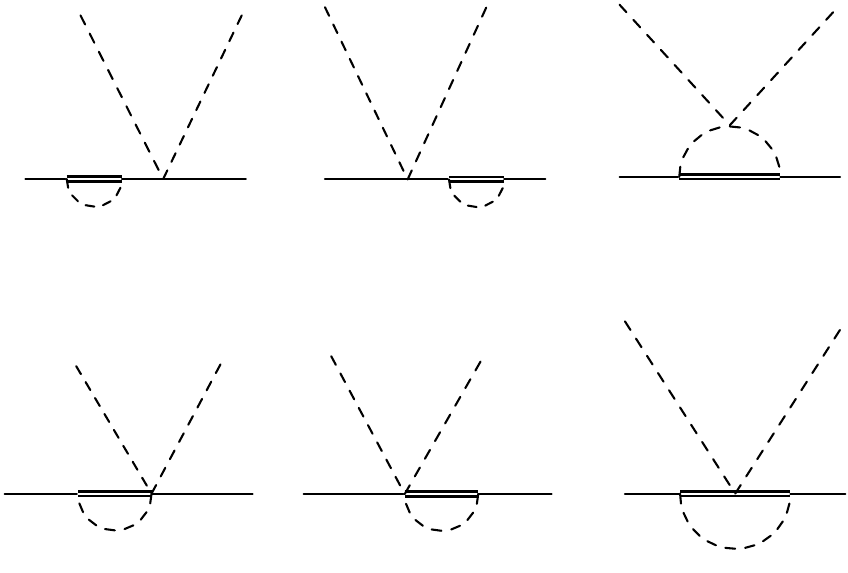}}\label{fig1b}
 }
 \hspace{\stretch{0.5}}
\caption{Nonvanishing loop diagrams for the pseudoscalar meson and
charmed meson scattering lengths to $O(\epsilon^3)$ with
HB$\chi$PT. The dashed lines represent the pseudoscalar Goldstone
bosons. Both the thin solid lines and thick solid lines represent
charmed baryons. The internal thin solid lines represent the
charmed baryons in the same representation as the external baryons
while the internal thick solid lines represent all possible
charmed baryons. } \label{figLoopDiag}
\end{figure}

We can get the scattering length to $O(\epsilon^3)$ by calculating
the corresponding tree diagrams and loop diagrams in Fig.
\ref{figLoopDiag}. The diagrams in Fig. \ref{fig1a} are made up of
the Weinberg-Tomozawa vertices, while the diagrams in Fig.
\ref{fig1b} are made up of two axial coupling vertices. We use the
dimensional regularization and modified minimal subtraction to
deal with the divergence generated by loop diagrams. All the
divergence will be canceled after the redefinition of the low
energy constants at $O(\epsilon^3)$ and renormalization of the
wavefunction \cite{Liu2012b}.

With the strict isospin symmetry, there are eleven independent
channels for $\phi B_{\bar 3}$ scattering, and nineteen
independent channels for either $\phi B_{6}$ or $\phi B_{6}^*$
scattering. We can express the scattering length order by order.
For example, the scattering length $a^{(2)}_{\pi\Sigma_c}$ in the
$\pi\Sigma_c$ channel with the isospin 2 is
\begin{eqnarray}
a^{(2)}_{
\pi\Sigma_c}&=&\frac{1}{4\pi(1+m_\pi/m_{\Sigma_c})}\left\{-\frac{m_\pi}{f_\pi^2}\right\}
               +\frac{1}{4\pi(1+m_\pi/m_{\Sigma_c})}\left\{-\frac{m_\pi^2
                       (24 c_0+4 c_1-3 c_2-6 c_4)}{6 f_\pi^2}-\frac{m_\pi^2(g_1^2+2g_2^2)}{8M_0 f_\pi^2}\right\}
                       \nonumber\\&&
               +\frac{1}{4\pi(1+m_\pi/m_{\Sigma_c})}\left\{
[-\frac{m_\pi^3 (g_1^2+2g_2^2) }{16 f_\pi^2 M_0^2}+\frac{m_\pi^2
\delta_2(g_1^2+g_2^2)}{8 f_\pi^2 M_0^2}]
\right.\nonumber\\&&\qquad +[-\frac{1}{8}
V\left(m_K^2,-m_\pi\right) \left. -\frac{1}{2}
V\left(m_\pi^2,-m_\pi\right)-\frac{1}{4}
V\left(m_\pi^2,m_\pi\right)-\frac{1}{9} m_\pi^2
W_2(m_\eta)+m_\pi^2 W_2(m_\pi)] \right\},
\end{eqnarray}
where $M_0$ is the mean mass of charmed triplet baryons, and $c_i$
is the LEC at $O(\epsilon^2)$. The scalar functions $V$ and $W_2$
come from the loop correction, and their definitions can be found
in Ref. \cite{Liu2012b}.

To get the final numerical results, we should determine the LECs
of Lagrangians. One obtains $|g_2|=0.60$ and $|g_4|=1.0$ by
fitting the decay widths of $\Sigma_c$ and $\Sigma_c^*$
\cite{Meguro2011}. Due to lack of the experimental data, we cannot
fit all the LECs. One can get $|g_1|=\sqrt{8/3}|g_2|$ with the
quark model symmetry, and $|g_3|=\sqrt3/2 |g_1|$, $|g_5|=3/2
|g_1|$, $|g_6|=0$ with heavy quark spin symmetry. We use the SU(4)
flavor approximation to roughly determine the LECs at
$O(\epsilon^2)$,
\begin{equation}
  c_0=-0.79{~\rm GeV},\quad
  c_1=-0.98{~\rm GeV},\quad
  c_2=-2.07{~\rm GeV}-2\frac{\alpha'}{4\pi f},\quad
  c_3=\frac{\alpha'}{4\pi f},\quad
  c_4=-0.84{~\rm GeV},
\end{equation}
where we take the dimensionless constant $\alpha'$ to be in the
range of [-1,1] as in Ref. \cite{Guo2009}. As for the LECs at
$O(\epsilon^3)$, we use two different kinds of approximations. We
first assume that the LECs at $O(\epsilon^3)$ are zero
(Approximation I). We also let the LECs absorb the analytical
corrections from loop diagrams totally (Approximation II). The
analytical corrections are the polynomials of the momentum or mass
of the charmed baryons and light pseudoscalar mesons.

\section{Results and discussions}\label{secRsl}

Now we can get the scattering lengths of 49 independent channels
to the third order \cite{Liu2012b}. From our numerical results, we
notice that there exists attractive interaction in the following
channels with the positive real parts of scattering lengths: ${
\pi\Xi_c}^{(1/2)}$, ${ K\Xi_c}^{(0)}$, ${ K\Xi_c}^{(1)}$, ${ \bar
K\Lambda_c}^{(1/2)}$, ${ \bar K\Xi_c}^{(0)}$, ${
\eta\Lambda_c}^{(0)}$, ${ \eta\Xi_c}^{(1/2)}$ , ${
\pi\Xi^\prime_c}^{(1/2)}$, ${ \pi\Sigma_c}^{(0)}$, ${
\pi\Sigma_c}^{(1)}$, ${ K\Omega_c}^{(1/2)}$, ${
K\Xi^\prime_c}^{(0)}$, ${ K\Xi^\prime_c}^{(1)}$, ${
K\Sigma_c}^{(1/2)}$, ${ \bar K\Xi^\prime_c}^{(0)}$, ${ \bar
K\Sigma_c}^{(1/2)}$, ${ \bar K\Sigma_c}^{(3/2)}$, ${
\eta\Omega_c}^{(0)}$, ${ \eta\Xi^\prime_c}^{(1/2)}$, ${
\eta\Sigma_c}^{(1)}$, ${ \pi\Xi^*_c}^{(1/2)}$, ${
\pi\Sigma^*_c}^{(0)}$, ${ \pi\Sigma^*_c}^{(1)}$, ${
K\Omega^*_c}^{(1/2)}$, ${ K\Xi^*_c}^{(0)}$, ${ K\Xi^*_c}^{(1)}$,
${ K\Sigma^*_c}^{(1/2)}$, ${ \bar K\Xi^*_c}^{(0)}$, ${ \bar
K\Sigma^*_c}^{(1/2)}$, ${ \bar K\Sigma^*_c}^{(3/2)}$, ${
\eta\Omega^*_c}^{(0)}$, ${ \eta\Xi^*_c}^{(1/2)}$, and ${
\eta\Sigma^*_c}^{(1)}$, where the superscripts in parentheses
refer to the isospin.

There is no great difference of the scattering lengths between the
two different approximations of the LECs at $O(\epsilon^3)$ in
most of the channels. There are 41 channels, in each of which the
difference between the different approximations is at most 20\% of
the smaller scattering length. The convergence of the chiral
expansion with Approximation II is better than that with
Approximation I generally, especially in the following channels ${
\bar K\Lambda_c}^{(1/2)}$, ${ K\Omega_c}^{(1/2)}$, ${
K\Sigma_c}^{(3/2)}$, ${ \bar K\Omega_c}^{(1/2)}$, ${ \bar
K\Sigma_c}^{(1/2)}$, ${ K\Omega^*_c}^{(1/2)}$, ${
K\Sigma^*_c}^{(3/2)}$, ${ \bar K\Omega^*_c}^{(1/2)}$, and ${ \bar
K\Sigma^*_c}^{(1/2)}$.

We investigate the effect of the mass difference $\delta_i$
between charmed baryons. The mass difference $\delta_i$ only
affects the scattering lengths at $O(\epsilon^3)$. In Table.
\ref{tabCom}, we list the $\phi B_{\bar 3}$ scattering lengths at
$O(\epsilon^3)$ with $\delta_i$ in Eq. (\ref{eqMssDif}) and with
$\delta_i=0$. From the table, we notice that the difference with
different $\delta_i$ is at most 10\% of the smaller scattering
length in seven channels among the total ten channels. The effect
of the mass difference for the $\phi B_{6}$ and $\phi B_{6}^*$
scattering lengths is small in most of the channels similar to
that for the $\phi B_{\bar 3}$ scattering lengths.

\begin{table}[htbp]
\begin{center}
\begin{tabular}{c|cc|cc}
\hline
                                  &\multicolumn{2}{|c|}{$\delta_1=67, ~  \delta_2=127, ~  \delta_3=194 {~\rm (MeV)}$}   &\multicolumn{2}{c}{$\delta_1=\delta_2=\delta_3=0 $}\\
~                                 &Approximation I      &Approximation II&Approximation I      &Approximation II \\
\hline
$a^{(1)}_{ \pi\Lambda_c}$         &-0.008               &-0.017          &-0.02                &-0.02          \\
$a^{(1/2)}_{ \pi\Xi_c}$           &0.011                &0.00044         &0.0079               &-0.0027        \\
$a^{(3/2)}_{ \pi\Xi_c}$           &-0.048               &-0.043          &-0.051               &-0.046         \\
\hline
$a^{(1/2)}_{ K\Lambda_c}$         &-0.24                &-0.15           &-0.24                &-0.15          \\
$a^{(0)}_{ K\Xi_c}$               &0.052                &-0.090          &0.095                &-0.091          \\
$a^{(1)}_{ K\Xi_c}$               &0.27+0.18$i$         &0.23+0.18$i$    &0.25+0.18$i$         &0.25+0.18$i$   \\
$a^{(1/2)}_{ \bar K\Lambda_c}$    &0.084+0.27$i$        &-0.0071+0.27$i$ &0.084+0.27$i$        &-0.0071+0.27$i$\\
$a^{(0)}_{ \bar K\Xi_c}$          &0.93                 &0.75            &0.88                 &0.78           \\
$a^{(1)}_{ \bar K\Xi_c}$          &-0.24                &-0.15           &-0.23                &-0.14          \\
\hline
$a^{(0)}_{ \eta\Lambda_c}$        &0.095+0.19$i$        &0.083+0.19$i$   &0.088+0.19$i$        &0.088+0.19$i$  \\
$a^{(1/2)}_{ \eta\Xi_c}$          &0.036+0.098$i$       &0.028+0.098$i$  &0.026+0.098$i$       &0.026+0.098$i$ \\
\hline\hline
\end{tabular}
\caption{the $\phi B_{\bar 3}$ scattering lengths at
$O(\epsilon^3)$ with different mass difference $\delta_i$ in units
of fm}\label{tabCom}
\end{center}
\end{table}

In summary, we have calculated the pseudoscalar meson and charmed
baryon scattering lengths to $O(\epsilon^3)$ with HB$\chi$PT.
There exists attractive interaction in some channels, which might
lead to the formation of some loosely bound molecules. We notice
that the convergence becomes better in most of the channels with
the non-analytic approximation. The effect of the mass difference
between charmed baryons is not large. Hopefully our results will
be useful for the chiral extrapolation of lattice QCD simulation
of the pion heavy baryon strong interaction.

\section{Acknowledgments}
This work is supported by the National Natural Science Foundation
of China under Grants 11075004, 11021092 and Ministry of Science
and Technology of China(2009CB825200).

\end{document}